\documentclass[twoside,twocolumn,nofootinbib,english]{revtex4-1}
\usepackage{mathptmx}
\usepackage[T1]{fontenc}
\usepackage[latin9]{inputenc}
\usepackage{xcolor}
\usepackage{pdfcolmk}
\usepackage{units}
\usepackage{amssymb}
\usepackage{graphicx}
\usepackage{hyperref}
\PassOptionsToPackage{normalem}{ulem}
\usepackage{ulem}
\def\be{\begin{equation}}
\def\ee{\end{equation}}
\def\bea{\begin{eqnarray}}
\def\eea{\end{eqnarray}}


\usepackage{babel}
\begin{document}

\title{Zero-site DMRG and the optimal low-rank correction}

\author{Yuriel Núñez-Fernández, Gonzalo Torroba}

\affiliation{Centro Atómico Bariloche and Instituto Balseiro, CONICET, R8402AGP,  Argentina.}
\begin{abstract}

A zero-site density matrix renormalization algorithm (DMRG0) is proposed
to minimize the energy of matrix product states (MPS). Instead of
the site tensors themselves, we propose to optimize sequentially the ``message'' tensors between neighbor sites, which contain the singular values of the bipartition. This
leads to a local minimization step that is independent of the physical dimension of the site. Conceptually, it separates the optimization and decimation steps in DMRG. Furthermore, we introduce two new global perturbations
based on the optimal low-rank correction to the current state, which are used to avoid local minima. They are determined variationally as the MPS closest to
the one-step correction of the Lanczos or Jacobi-Davidson eigensolver,
respectively. These perturbations mainly decrease the
energy and are free of hand-tuned parameters. Compared to existing single-site enrichment proposals, our approach
gives
similar convergence ratios per sweep while the computations are cheaper
by construction. 
Our methods may be useful in systems with many physical degrees of freedom per lattice site. We test our approach on the periodic Heisenberg
spin chain for various spins, and on free electrons on the lattice.

\end{abstract}
\maketitle

\section{Introduction}\label{sec:intro}

The density matrix renormalization group (DMRG), introduced by White~\cite{white1992density, white1993density}, provides a very powerful approach to study quantum systems in one dimension. The success of the method is based on the fact that its variational wavefunction efficiently captures the local entanglement properties of ground states. Besides static properties, it has been extended to study excited states, dynamics, etc.; see~\cite{schollwock2011density} for a recent review. Nevertheless, key challenges need to be resolved for a broader applicability of DMRG, most notably for quantum systems in higher dimensions. Here, the area law of entanglement makes the algorithm exponentially costly. Therefore, improving the DMRG approach remains an important goal, both theoretically and for applications.

\begin{figure}[h!]
\centering
\includegraphics[width=1.\linewidth]{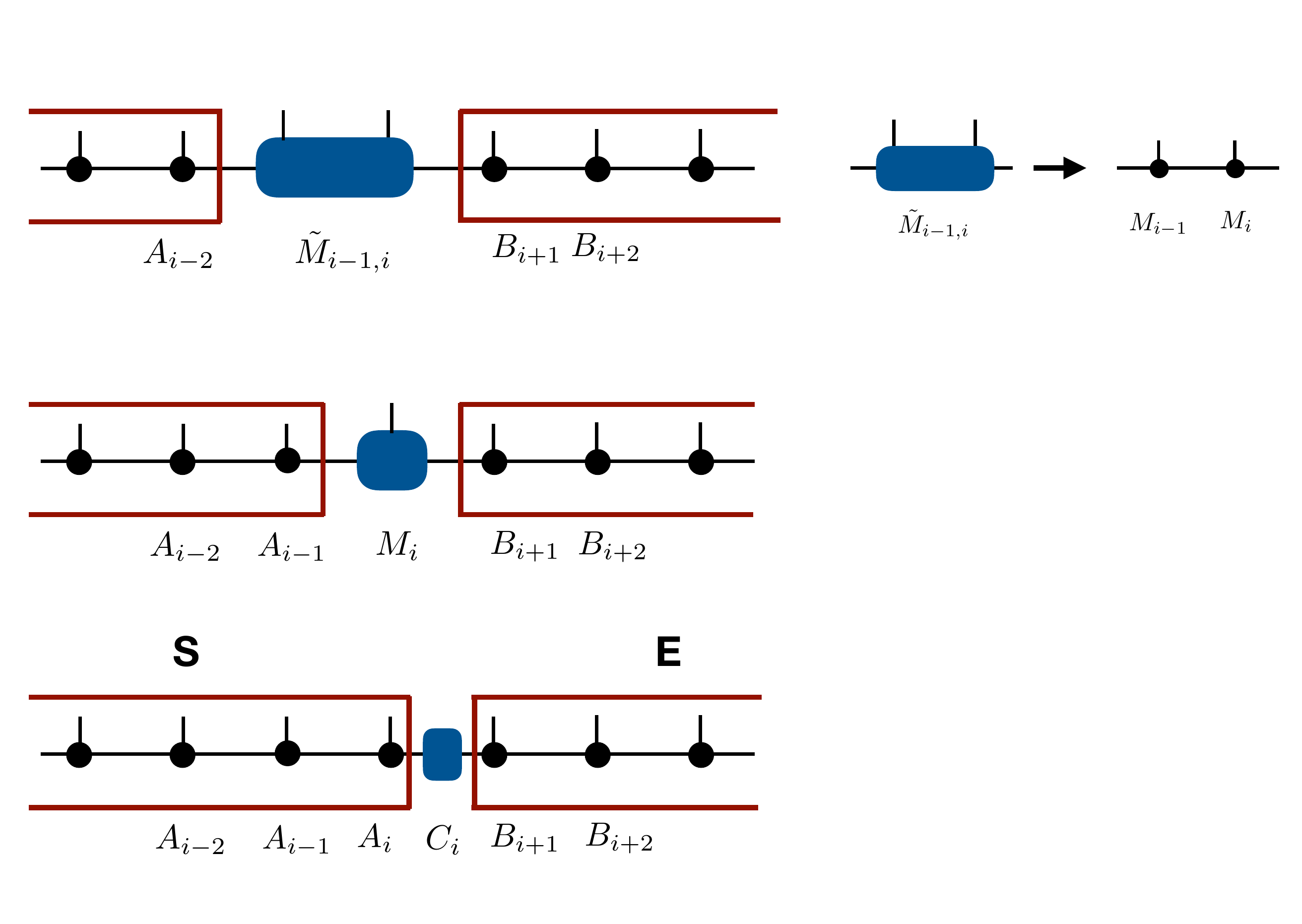}
\caption{\small{MPS representation of DMRG. Black circles represent the tensors in the ansatz (\ref{eq:MPS1}); a vertical line corresponds to a physical index, while horizontal lines are bond indices. A line connecting two dots means index-contraction. The object in the blue box is being optimized. We choose a canonical representation, where matrices $A_i$ to the left of the optimized object are left-normalized; similarly matrices $B_j$ to the right are right-normalized.
 White's two-site algorithm is shown in the upper panel, and the single site DMRG is displayed in the middle panel. Our work introduces the 0-side DMRG, showed here in the lower panel. At step $i$, we optimize the $m \times m$ matrix $C_i$ that contains the singular values of the bipartition.}}\label{fig:mps-dmrg}
\end{figure}

A key conceptual development was the realization that the DMRG ansatz gives a ground state of matrix product state (MPS) form~\cite{PhysRevLett.75.3537, PhysRevB.55.2164}
\be\label{eq:MPS1}
|\psi\rangle=\sum_{\sigma_{1}...\sigma_{L}}M_{1}^{\sigma_{1}}...M_{L}^{\sigma_{L}}|\sigma_{1}...\sigma_{L}\rangle\,.
\ee
Here, $\sigma_{i}$ (of dimension
$d_{i}$ with $i=1,2,...,L$) labels the physical degree of freedom
of the site $i$ for a system of $L$ sites. $M_{i}$ is a rank-3
tensor of dimensions $m_{i-1}\times d_{i}\times m_{i}$. In this language, the variational optimization is carried over objects formed out of the tensors $M_i$ (see below), while the decimation involves restricting the dimensions to some fixed bond dimension $m$, namely $m_i \le m$. Keeping a fixed upper bound $m$ on the bond dimension of each $M_i$ entails an exponential compression of the basis of states. 
The reformulation of the DMRG in terms of MPS has led to new applications and improvements~\cite{doi:10.1063/1.4955108}, as well as placing the DMRG in the more general context of tensor networks~\cite{ORUS2014117}. This is the language we will use here; the graphical representation of the state is explained in Fig.~\ref{fig:mps-dmrg}.

The DMRG algorithm involves four main steps: \emph{1)} a choice of basic object or block to optimize; \emph{2)} the optimization procedure; \emph{3)} an enrichment in order to avoid local minima; and \emph{4)} the decimation step. In MPS terms, White's approach corresponds to a \emph{2-site} DMRG: the object being optimized has two physical indices. At a given step, this is shown in the upper part of Fig.~\ref{fig:mps-dmrg}, with the basic block denoted by $\tilde M$. This quantity has two physical indices and hence lives on a space bigger than the basic MPS tensors. This leads to an enrichment of the variational space. After the optimization, the 2-site object $\tilde M$ is re-expressed in terms of the MPS tensors $M_i$, as shown in the right part of the figure.

From this perspective, a natural modification is to optimize directly the tensors $M_i$, and this leads to the \emph{single-site} DMRG~\cite{white2005density}. We illustrate this method in the middle panel of Fig.~\ref{fig:mps-dmrg}. Changing the optimization from 2-site blocks to a single-site block decreases the dimension of the variational matrices from $md \times md$ to $md \times m$, and is expected to decrease the optimization time by a factor of order $d$. Two aspects are worth emphasizing. First, the extra site in the 2-site DMRG played an important role in enriching the variational space, so the single-site DMRG needs an independent enrichment step. This is required to make sure that the most relevant states are present in the reduced density matrix, and hence avoid local minima. The other point is that only $m$ out of the $md$ rows of $M_i$ can be linearly independent. As a result, the optimization step entails a decimation.

In this work we propose the \emph{0-site} DMRG (DMRG0), where the $m \times m$ singular-value matrices $C_i$ (obtained by decomposing $M_i^{\sigma_i} = A_i^{\sigma_i} C_i$) are optimized. See the lower panel in Fig.~\ref{fig:mps-dmrg}. Implementing the 0-site DMRG is important for various reasons.
Conceptually, it formulates the optimization of the wavefunction
(which dominates the costs in DMRG) considering only a pure bipartition
at a time, that is, asking for the optimal wavefunction expressed
in terms of the system $S$ and the environment $E$ renormalized basis (both of size $m$). In contrast, current computational schemes 
\cite{white2005density, schollwock2005dmrg,hallberg2006newtrends,schollwock2011density,chan2011density, dolgov2015corrected, hubig2015strictly}
explicitly include one
or two sites as part of $S$ and/or $E$ when the local optimization
step is performed. Thus the system becomes tripartite at this step. Another aspect is that we now expect the local optimization cost to be independent of the physical dimension $d$, as opposed to the single and 2-site algorithms, where the cost depends explicitly on $d$. This is a very attractive feature for the simulation of systems with a large number of degrees of freedom per site.\footnote{It is worth emphasizing that we are referring here to the local optimization cost. The independence with $d$ does not eliminate the more fundamental limitation that the amount of entanglement may just be too large to simulate with a bounded bond dimension, as occurs for instance in gapless systems.} Another theoretical advantage is that the optimization and decimation steps are clearly distinct in DMRG0: the optimization occurs over a full-rank $m$ metric $C_i$. There is no redundant information here, as opposed to the single-site algorithm that optimizes over $md \times m$ matrices. Finally, we mention that DMRG0 may be interesting for tangent-space methods~\cite{PhysRevB.94.165116}, where the message matrix $C_i$ and its optimization appear naturally.

However, an important obstacle to this approach could be that the reduced variational space of the $C_i$ may not be rich enough to include important fluctuations between the system and the environment. This can exacerbate the problem of local minima. A successful realization of DMRG0 necessarily needs to face this challenge. An important part of this work will then be to solve this in an efficient manner. This leads us to introduce an
enrichment step based on the optimal low-rank correction to the global
state. We will show that this method markedly increases the convergence of
the algorithm and avoids metastable solutions. We will compare it with existing enrichment methods, finding various advantages related to global convergence properties and the absence of arbitrary external parameters that need to be tuned during the enrichment step.

The goal of this work is to describe and implement the DMRG0 algorithm.
The outline of the paper is as follows. In Sec.~\ref{sec:0dmrg} we present the zero-site DMRG and explain its basic properties. In Sec.~\ref{sec:optimal} we introduce the enrichment step, with two approximate schemes (Lanczos and Jacobi-Davidson) to obtain the optimal low-rank correction. We also review the previous approaches~\cite{white2005density, dolgov2015corrected, hubig2015strictly}, and establish the equivalence between~\cite{white2005density} and~\cite{ hubig2015strictly} . Sec.~\ref{sec:Algorithm} summarizes our algorithm. Sec.~\ref{sec:results} presents numerical results for the Heisenberg spin chain with spins $S=1$, $S=3$ and $S=5$, and for free fermions, and compare with existing single-site enrichment methods. Sec.~\ref{sec:concl} contains our conclusions and perspectives.

\section{Zero-site DMRG}\label{sec:0dmrg}

A quantum state $\psi$ of the form (\ref{eq:MPS1}) is called 
a matrix product state (MPS); see \cite{schollwock2011density} for a detailed review. While the representation is exact for sufficiently large bond dimensions $m_i$,
in practice
$m_{i}\le m$. This is a key part of the decimation or renormalization of the relevant degrees of freedom. 

An important property of the MPS is its gauge degree of freedom. For
arbitrary invertible matrices $X_{i}$, the identity $I_{i}=X_{i}^{-1}X_{i}$
can be inserted between $M_{i}$ and $M_{i+1}$, effectively changing
the matrices to $\tilde{M}_{i}^{\sigma}=X_{i-1}M_{i}^{\sigma}X_{i}^{-1}$
while keeping the state invariant. It allows us to choose the so-called
left (right) normalization for the matrices $\tilde{M}_{i}^{\sigma}=A_{i}^{\sigma}$
($\tilde{M}_{i}^{\sigma}=B_{i}^{\sigma}$) satisfying
\be
\sum_{\sigma}\left(A_{i}^{\sigma}\right)^{\dagger}A_{i}^{\sigma}=I_{i}\;\mbox{ or }\;\sum_{\sigma}B_{i}^{\sigma}\left(B_{i}^{\sigma}\right)^{\dagger}=I_{i-1}\,.
\ee
We now introduce (as in \cite{haegeman2016unifying}) the MPS \emph{zero-site}
\emph{canonical} form at site $i$:
\begin{equation}
|\psi\rangle=\sum_{\sigma_{1}...\sigma_{L}}A_{1}^{\sigma_{1}}...A_{i}^{\sigma_{i}}C_{i}B_{i+1}^{\sigma_{i+1}}...B_{L}^{\sigma_{L}}|\sigma_{1}...\sigma_{L}\rangle\mbox{, }\label{eq:mps_cano}
\end{equation}
similar to the single-site canonical form where the central
matrix $M_{i}^{\sigma_{i}}$ is decomposed as 
\be
M_{i}^{\sigma_{i}}=A_{i}^{\sigma_{i}}C_{i}\,.
\ee 
The matrix $A_i$ is left-normalized, and $C_i$ contains the singular values of $M_i$. Note that $C_i$ does not contain the physical index $\sigma_i$ and its dimension $d_i$ -- we associate it to the link between $A_i$ and $B_{i+1}$.
We illustrate this in the lower panel of Fig.~\ref{fig:mps-dmrg}.
One advantage of this representation is its local expression for
the square norm $\langle\psi|\psi\rangle=\mbox{tr}(C_{i}^{\dagger}C_{i})$.
We refer to $C_{i}$ as the ``message'' between site
tensors -- it contains the singular values and the entanglement of the
bipartition in this case. In DMRG terminology, the products
$A_{1}^{\sigma_{1}}...A_{i}^{\sigma_{i}}$ and $B_{i+1}^{\sigma_{i+1}}...B_{L}^{\sigma_{L}}$
represent the left and right renormalized basis for $S$ (system) and $E$ (environment)
respectively, and $C_{i}$ is the (strictly bipartite) wavefunction.

If the Hamiltonian is also written as a matrix product operator (MPO),
\be
\hat{H}=\sum_{\sigma_{1}...\sigma_{L}}W_{1}^{\bar{\sigma}_{1}\sigma_{1}}...W_{L}^{\bar{\sigma}_{L}\sigma_{L}}|\bar{\sigma}_{1}...\bar{\sigma}_{L}\rangle\langle\sigma_{1}...\sigma_{L}|\mbox{,}
\ee
then the energy of a normalized state $\psi$ is 
\begin{equation}
E=\langle\psi|\hat{H}|\psi\rangle=C_{i}^{\dagger}\langle\partial_{i}\psi|\hat{H}|\partial_{i}\psi\rangle C_{i}\equiv C_{i}^{\dagger}H_{\psi\psi}^{i}C_{i}\mbox{,}\label{eq:energy}
\end{equation}
where $|\partial_{i}\psi\rangle$ is the derivative of $|\psi\rangle$
with respect to $C_{i}$ in (\ref{eq:mps_cano}). The effective operator
$H_{\psi\psi}^{i}=\langle\partial_{i}\psi|\hat{H}|\partial_{i}\psi\rangle$
does not depend on $C_{i}$, and can be calculated recursively using
the transfer matrices for $\hat{H}$,
\begin{equation}
H_{\psi\psi}^{i}=L_{i}R_{i+1}\,.\label{eq:Heff}
\end{equation}
Here, $L_{i}=L_{i-1}T_{i}^{A}$, $R_{i}=T_{i}^{B}R_{i+1}$ with $L_{0}=R_{L+1}=1$,
$T_{i}^{A}$ is the left transfer matrix $T_{i}^{A}=\left(A_{i}^{\bar{\sigma}}\right)^{\dagger}W_{i}^{\bar{\sigma}\sigma}A_{i}^{\sigma}$, and $T_{i}^{B}$ is the right transfer matrix $T_{i}^{B}=\left(B_{i}^{\bar{\sigma}}\right)^{\dagger}W_{i}^{\bar{\sigma}\sigma}B_{i}^{\sigma}$, for $i=1,2,...,L$.

Our zero-site DMRG proposal (DMRG0) is to optimize one tensor $C_i$ at a time. In a DMRG step, the position $i$ and the renormalized operators $L_{i}$,
$R_{i+1}$ are fixed while the wavefunction $C_{i}$ is updated. $E$
and $C_{i}$ are the lowest eigenvalue and eigenvector, respectively,
of the effective Hamiltonian $H_{\psi\psi}^{i}$. An
iterative eigensolver like Lanczos is used to diagonalize (\ref{eq:Heff})
starting from the previous $C_{i}$ until a given tolerance is reached.
The position $i$ is then changed to $i+1$ in (\ref{eq:mps_cano}),
performing a matrix decomposition of $M_{i+1}^{\sigma}=C_{i}B_{i+1}^{\sigma}=A_{i+1}^{\sigma}C_{i+1}$.
An analogous step is performed for the change from $i$ to $i-1$.

As discussed in Sec.~\ref{sec:intro}, the main problem to solve in this approach is how to avoid local minima. To this end, we will now present a new enrichment method based on the optimal low-rank correction.

\section{Enrichment via optimal low-rank correction}\label{sec:optimal}

The MPS ansatz is highly non-linear in its parameters, the matrices
$M_{i}^{\sigma}$. Despite the success of the DMRG proposal to
optimize one tensor at a time, there is the danger of being trapped in
local minima, especially for single-site effective wavefunction approaches.
For our DMRG0, in principle we expect an even worse situation; we will analyze this in examples below in Sec.~\ref{sec:results}. The development of an efficient space enrichment method is then central to the success of DMRG0.

The space enrichment methods ~\cite{white2005density, dolgov2015corrected, hubig2015strictly} are local, i.e. they enrich only one site-tensor at a time. They are based on the application of renormalized operators
living on $S$ (of dimension $md$) to the single-site wavefunction
of dimensions $md\times m$. This introduces the possibility that the
renormalization from $md$ to $m$ changes the wavefunction. We review these approaches in Sec.~\ref{subsec:previous}, establishing the formal equivalence between ~\cite{white2005density} and \cite{hubig2015strictly}.
These ideas, however,
are not directly applicable to our DMRG0 because the effective wavefunction
$C_{i}$ is a full-rank $m\times m$ matrix with entries on $S$ and
$E$ basis, both of size $m$. In Sec.~\ref{subsec:optimal} we present our new global proposal for the optimal correction, and two approximate schemes for obtaining it.

\subsection{Previous approaches and equivalence}\label{subsec:previous}

To simplify the explanation, let us focus on the decimation step in
DMRG. Given a bipartition $\{S,E\}$ equipped with their respective
basis, a state $|\psi\rangle$ corresponds to a matrix $M$. The reduced
density matrix for $S$ is $\rho=MM^{\dagger}$. In the original DMRG
\cite{white1992density,white1993density}, $\rho$ is diagonalized
$\rho=UDU^{\dagger}$, its eigenvalues $D$ are truncated to $\bar{D}$
containing the $m$ largest values, and $U$ is truncated to $\bar{U}$
containing the corresponding $m$ eigenvectors, that is, $\rho\approx\bar{U}\bar{D}\bar{U}^{\dagger}$.
In this approximation, the operators in $S$ are renormalized according
to $\bar{O}=\bar{U}^{\dagger}O\bar{U}$, and the states according
to $\bar{v}=\bar{U}^{\dagger}v$. In particular, $|\psi\rangle$ transforms
as $\bar{M}=\bar{U}^{\dagger}M$.

The first approach to enrich the space was introduced by S. White in~\cite{white2005density}. Anticipating later incorporations of
relevant states in the environment basis, the density matrix of $S$ is perturbed using the Hamiltonian terms living in $S$,
\begin{equation}
\tilde{\rho}=\rho+\beta^{2}\sum_{\gamma}L_{\gamma}\rho\left(L_{\gamma}\right)^{\dagger}\,.\label{eq:DMP}
\end{equation}
This replaces $\rho$ by $\tilde \rho$ in the decimation step above. Here,
the number $\beta^{2}$ is a small weight tuned by hand, and $L_{\gamma}$
($R_{\gamma}$) are the renormalized operators of $S$ ($E$) appearing
in the Hamiltonian $\hat{H}=\sum_{\gamma}\hat{L}_{\gamma}\otimes\hat{R}_{\gamma}$
for the given bipartition $\{S,E\}$. The resulting density matrix $\tilde \rho$ is renormalized to keep the largest $m$ eigenvalues.

The second approach \cite{dolgov2015corrected}, as part of the alternating minimal energy (AMEn) algorithm, enriches the space by directly enlarging the wavefunction
$M$ to $\tilde{M}=\left(\begin{array}{cc}
M & P\end{array}\right)$. The technique is called subspace expansion. Starting from the innocuous transformation (adapted to our notation):
\be\label{eq:Mlarge}
M=\underbrace{\left(\begin{array}{cc}
M & P\end{array}\right)}_{\tilde{M}}\cdot\left(\begin{array}{c}
I\\
\mathbf{0}
\end{array}\right)\mbox{,}
\ee
where $I$, $\mathbf{0}$ are the appropriate identity and null matrices,
respectively, Ref.~\cite{dolgov2015corrected} uses $\tilde{M}$ to grow the subsystem $S$ basis. In the next step, this allows to choose a richer state by changing the initially vanishing components in (\ref{eq:Mlarge}). Choosing $P$ as the single-site wavefunction of the (aproximate) residual $\left(H- E \right) |\psi\rangle $, with $E=\langle \psi| H |\psi\rangle$, this method guarantees convergence to the global minima~\citep{dolgov2014alternating}.

The third approach~\cite{hubig2015strictly}, as part of the DMRG3S algorithm, also uses the subspace expansion technique, and is based on a perturbation of the form
\begin{equation}
P=\beta\left(\begin{array}{cccc}
L_{1}M & L_{2}M & ... & L_{\gamma}M\end{array}\right)\,.\label{eq:sub_exp}
\end{equation}
It makes a singular value decomposition
(SVD) $\tilde{M}=U_{2}sV^{\dagger}$, followed by a truncation $\tilde{M}\approx\bar{U}_{2}\bar{s}\bar{V}^{\dagger}$
containing the largest $m$ singular values of $s$. Because some
reordering of $\tilde{M}$ takes place during this truncation, the original
state is modified/enriched. The basis of $S$ is rotated
with $\bar{U}_{2}$ and the new state is
\be
\bar{N}=\bar{s}\bar{V}^{\dagger}\cdot\left(\begin{array}{c}
I\\
\mathbf{0}
\end{array}\right).
\ee
Results comparable to~\cite{white2005density} are obtained, at a lower computational
cost. Note that $\beta$ in (\ref{eq:sub_exp}) assigns the same weight to all the states used for enrichment; this is not necessarily the optimal choice~\citep{dolgov2015corrected}.

Let us discuss the connection between~\cite{white2005density} and~\cite{hubig2015strictly}. For this, we note that the density matrix calculated from $\tilde M$ given (\ref{eq:sub_exp}) is the same as (\ref{eq:DMP}). Therefore the enrichment steps are equivalent in the $S$ subsystem. In more detail, since
$\tilde{\rho}=\tilde{M}\cdot\tilde{M}^{\dagger}$,
the diagonalization of $\tilde{\rho}=U_2DU_2^{\dagger}$ can be extracted
from the SVD of $\tilde{M}=U_2 sV^{\dagger}$ where $D=s^{2}$. The state
$\bar{M}$ obtained after truncation using $\tilde{\rho}\approx\bar{U_2}^{\dagger}\bar{D}\bar{U_2}$
is
\begin{eqnarray}
\bar{M} & = & \bar{U_2}^{\dagger}\cdot M=\bar{U_2}^{\dagger}\left(\begin{array}{cc}
M & P\end{array}\right)\cdot\left(\begin{array}{c}
I\\
\mathbf{0}
\end{array}\right) \nonumber \\
 & = & \bar{U_2}^{\dagger}\tilde{M}\cdot\left(\begin{array}{c}
I\\
\mathbf{0}
\end{array}\right)\approx\bar{U_2}^{\dagger}\bar{U_2}\bar{s}\bar{V}^{\dagger}\cdot\left(\begin{array}{c}
I\\
\mathbf{0}
\end{array}\right) \nonumber \\
 & = & \bar{N}\mbox{,}
\end{eqnarray}
which means that both approaches are equivalent also for the wavefunction
within the truncation error. They differ, however, in the representation,
and this is responsible for the difference in computational time.

\subsection{Optimal correction}\label{subsec:optimal}

A a natural way to enlarge
our variational space while keeping the computation tractable is to add a new state $\tilde \psi$
\be\label{eq:enrich}
\alpha\psi+\beta\tilde{\psi}\;,\;\alpha,\beta\in\mathbb{R}\,,
\ee
where the current (normalized) state $\psi$ is fixed and $\tilde{\psi}$
is the perturbation. We enrich the space by adding the new
direction $\tilde{\psi}$, and determine the coefficients $\alpha,\beta$
by the Rayleigh-Ritz method. That is, the
pair $(\alpha,\beta)^{t}$ is the lowest eigenvector of the $2\times2$
matrix:
\begin{equation}
H_{ab}=\langle a|\hat{H}|b\rangle\;,\;a,b\in\{\psi,\tilde{\psi}\}\mbox{.}\label{eq:h_2x2mat}
\end{equation}
We note that the alternating linear scheme
``ALS($t+z$)'' method~\cite{dolgov2014alternating}, developed
to solve linear systems, also enriches the approximate solution $t$ by adding the residual $z$.

We propose to obtain the perturbation $\tilde{\psi}$ by extremizing the energy,
\begin{equation}
\langle\delta\tilde{\psi}|\left[\alpha P\hat{H}|\psi\rangle+\beta P(\hat{H}-\lambda)P|\tilde{\psi}\rangle\right]=0\mbox{,}\label{eq:lagrangian}
\end{equation}
Here $\lambda$ is a Lagrange multiplier coming from the normalization condition. The projector $P=1-|\psi\rangle\langle\psi|$,
satisfying $P\psi=0$ and $P\tilde{\psi}=\tilde{\psi}$, implements the orthogonality condition $\langle\tilde{\psi}|\psi\rangle=0$.

Eq.~(\ref{eq:lagrangian}) is the exact condition that determines the \emph{optimal correction}. It can be cast as an
inhomogeneous eigenvalue problem of the form $A\vec{x}=\lambda\vec{x}+\vec{b}$
for the Hermitian matrix $A=P\hat{H}P$, $\vec{b}=-P\hat{H}|\psi\rangle$,
$\vec{x}=\nicefrac{\beta}{\alpha}\ P|\tilde{\psi}\rangle$. See e.g.~\cite{Dongarra2000templates} for methods to solve such problems in linear algebra. 

We now present two 
approximate iterative schemes for solving (\ref{eq:lagrangian}); they are motivated by considering a small correction to
the current state $\psi$, but are more broadly applicable.
A small correction $\beta\ll\alpha$ gives rise to
a residual (or Lanczos) scheme, while approximating $\lambda\approx E=\langle\psi|\hat{H}|\psi\rangle$
obtains a Jacobi-Davidson scheme. Both perform very well, and their strengths and weaknesses are inherited from the respective original
methods. The Lanczos iteration is fast and straightforward
to implement as a direct computation; it is sensitive to the loss of orthogonality, it is problematic when there are
(exact or approximate) degenerate states, and it is hard to apply
to the interior eigenvalues due to the shift-and-invert mechanism
required. On the other hand, the Jacobi-Davidson iteration is determined
by a more involved inverse problem, which in turn must be solved
iteratively; it is sensitive to the use of preconditioning, but it is very powerful
even for degenerate states or interior eigenstates.

\subsubsection{\label{sub:Residual-correction}Lanczos correction}

If $\beta\ll\alpha=\sqrt{1-\beta^{2}}\approx1-\frac{1}{2}\beta^{2}$,
we approximate
\begin{equation}
\langle\delta\tilde{\psi}|\left[P\hat{H}|\psi\rangle-\lambda\beta P|\tilde{\psi}\rangle\right]=0\mbox{,}\label{eq:first_order_beta}
\end{equation}
recalling that at this stage $\lambda$
is unknown. Eq.~(\ref{eq:first_order_beta}) implies that
$\tilde{\psi}$ is parallel to $P\hat{H}|\psi\rangle$ ,
\begin{equation}
|\tilde{\psi}\rangle\propto\left[1-|\psi\rangle\langle\psi|\right]\hat{H}|\psi\rangle=\left(\hat{H}-E\right)|\psi\rangle.\label{eq:res}
\end{equation}
The perturbation is then determined by a global residual calculation (\ref{eq:res}), 
which is similar to both the ALS(t+z) and  the AMEn algorithms.

In our zero-site DMRG framework, numerical experiments show that
the above correction (\ref{eq:res}) implemented as a global correction
after each DMRG sweep works well, see for instance Fig. 1. We will choose $\tilde{m}=m$, but note that one can take even $\tilde{m}=2m$,
while keeping the global computational cost governed by the
optimization step. See Secs.~\ref{sec:Algorithm} and~\ref{sec:results} for more details.

The self-consistency of the residual (\ref{eq:res}) as an approximate
solution to the optimal correction (\ref{eq:lagrangian}) could be
checked by direct substitution. However, this is not of our concern here, since the goal is to enrich the space using some
well-motivated perturbation.

\subsubsection{Jacobi-Davidson correction}\label{sub:Jacobi-Davidson-correction}

Eventually, if the residual correction (\ref{eq:res}) becomes insufficient,
the following method can be applied. It can be verified that $\lambda$
represents the energy of the new state $\alpha\psi+\beta\tilde{\psi}$
when Eq.~(\ref{eq:lagrangian}) is solved exactly. Motivated by the
Jacobi-Davidson algorithm, we take $\lambda\approx E=\langle\psi|\hat{H}|\psi\rangle$
in (\ref{eq:lagrangian}), obtaining
\begin{equation}
|\tilde{\psi}\rangle\propto-\left[P(\hat{H}-\lambda)P\right]^{-1}P\hat{H}|\psi\rangle\mbox{.}\label{eq:Jacobi-Davidson}
\end{equation}
We keep the symbol $\lambda$ because in our calculation scheme some
energy better than $E$ is usually available. The approximate solution
of a linear system like (\ref{eq:Jacobi-Davidson}) has a well established
algorithm in the DMRG community, see for instance the calculation
of Green's function response of~\cite{kuhner1999dynamical, cirac2009variational,chan2017time}. In our case, some additional
remarks concerning the presence of the projector $P$ are needed.
Ignoring the normalization of $|\tilde{\psi}\rangle$, and retaking
the variational principle (\ref{eq:lagrangian}), we obtain
\begin{eqnarray*}
\langle\delta\tilde{\psi}|\hat{H}-\lambda|\tilde{\psi}\rangle-\langle\delta\tilde{\psi}|\hat{H}-\lambda|\psi\rangle\langle\psi|\tilde{\psi}\rangle\\
-\langle\delta\tilde{\psi}|\psi\rangle\langle\psi|\hat{H}-\lambda|\tilde{\psi}\rangle\\
+\langle\delta\tilde{\psi}|\psi\rangle(E-\lambda)\langle\psi|\tilde{\psi}\rangle & = & -\langle\delta\tilde{\psi}|P\hat{H}|\psi\rangle\mbox{.}
\end{eqnarray*}

As dictated by DMRG, we fix the canonical position $i$ to find one
tensor ($\tilde{C}_{i}$ of $|\tilde{\psi}\rangle$) at a time. In
this context, the relation $|\tilde{\psi}\rangle=|\partial_{i}\tilde{\psi}\rangle\tilde{C}_{i}$
implies $|\delta\tilde{\psi}\rangle=|\partial_{i}\tilde{\psi}\rangle\delta\tilde{C}_{i}$,
and we have the following equation for $\tilde{C}_{i}$
\begin{eqnarray}
\Big \lbrace H_{\tilde{\psi}\tilde{\psi}}^{i}-\left|c_{i}^{O}\right\rangle \left\langle c_{i}^{H}\right|-\left|c_{i}^{H}\right\rangle \left\langle c_{i}^{O}\right| \label{eq:Proj_JD}\\
+(E+\lambda)\left|c_{i}^{O}\right\rangle \left\langle c_{i}^{O}\right|-\lambda \Big \rbrace & \left|\tilde{c}_{i}\right\rangle = & \left|c_{i}^{H}\right\rangle -E\left|c_{i}^{O}\right\rangle \mbox{,}\nonumber 
\end{eqnarray}
 where $\left|x\right\rangle $ corresponds to the matrix $X$ treated
as a vector, $c_{i}^{H}=\langle\partial_{i}\tilde{\psi}|\hat{H}|\psi\rangle$,
and $c_{i}^{O}=\langle\partial_{i}\tilde{\psi}|\psi\rangle$.

An iterative algorithm can be used to solve (\ref{eq:Proj_JD}) starting
from the previous $\tilde{C}_{i}$. For instance, we can apply
the generalized minimal residual method (GMRES~\cite{Dongarra2000templates}), with computational cost similar to the Lanczos diagonalization
of (\ref{eq:Heff}). The Jacobi-Davidson correction appears
to be more expensive than the residual correction; it would be interesting to perform a comparison
of the convergence ratios of both methods.

\section{Algorithm}\label{sec:Algorithm}

We summarize the previous results in the following algorithm. 

Let us call a \emph{sweep} to the sequence of positions from $i=0$
to $i=L$ (sweeping right) followed by its reverse form $i=L$ down
to $i=0$ (sweeping left).
\begin{enumerate}
\item[(i)]  Initialize $\psi$ with random matrices and canonicalize them to
the position $i=0$. Set the current error $\epsilon=1$ (arbitrarily
large).
\item[(ii)]  Make a standard sweep for $\langle\psi|\hat{H}|\psi\rangle$ calculating
the ground state $\psi$ using tolerance $\sim0.1\epsilon$ to diagonalize
each $H_{\psi\psi}^{i}$ in Eq.~(\ref{eq:Heff}).
\item[(iii)]  Set $\tilde{\psi}=(H-E)\psi$; starting from the exact MPO-MPS product
apply the zip-up algorithm \cite{white2010mpomps} to compress $\tilde{\psi}$
to bond dimension $\tilde{m}$. 

\begin{enumerate}
\item If desired, an additional sweep setting $\tilde{\psi}$ as the Jacobi-Davidson
correction (\ref{eq:Jacobi-Davidson}), (\ref{eq:Proj_JD}) can be used.
The quantities $\langle\tilde{\psi}|\hat{H}|\tilde{\psi}\rangle$,
$\langle\tilde{\psi}|\psi\rangle$ should also be swept to solve
(\ref{eq:Proj_JD}).
\end{enumerate}
\item[(iv)]  Update $\psi$ using the compression of $\alpha\psi+\beta\tilde{\psi}$
to bond dimension $m$. Set $\epsilon=E-\lambda_{1}$ where $\lambda_{1}$
is the first eigenvalue of the $2\times2$ matrix $H$ in (\ref{eq:h_2x2mat}).
Since $\langle\tilde{\psi}|\psi\rangle\ne0$ the overlap matrix $O=\langle a|b\rangle$
with $a,b\in\{\psi,\tilde{\psi}\}$ should be taken into account,
yielding a generalized eigenvalue problem $H\vec{x}=\lambda O\vec{x}$
with Hermitian positive-definite $2\times2$ matrix $O$.
\end{enumerate}
Steps (ii-iv) are repeated until the energy $E$ (or $\lambda_{1}$)
does not change. Step (ii) is the
zero-site DMRG (DMRG0), while (iii-iv) are the basis enrichment
steps, in this case, based on the Lanczos (DMRG0-L) or Jacobi-Davidson
(DMRG0-JD) correction. Optionally, $n$ successive perturbations
can be applied; the respective algorithms are denoted by DMRG0-L$n$ and DMRG0-JD$n$.

As usual, the diagonalization (ii) is the most time-consuming part;
its cost per site scales as $O(2m^{3}wK)$, where $w$ is the MPO bond dimension
and $K$ is the number of eigensolver iterations. Notice the absence
of the physical site dimension $d$ in this cost. To control the number
$K$ we need both a good starting point (already provided) and we should avoid iterations far beyond the renormalization error of the
MPS. Step (iv) provides an appropriate error quantity $\epsilon$
to ask for during the diagonalization, keeping $K$ in the order of
few tens during the entire calculation. For comparison, the single-site
scheme scales as $O(2m^{3}wdK+d^{2}m^{2}w^{2}K)$ with a typical larger
value for $K$ because the local problem is $d$ times bigger.

Concerning our enrichment proposal, the cost of the Lanczos correction
is similar to that of the subspace expansion (\ref{eq:sub_exp}) in
DMRG3S. It is dominated by the SVD compression of a $mw\times md$
matrix, which scales as $O(m^{3}wd^{2})$. The cost of compressing the sum of two
MPSs is negligible $O(8m^{3}d)$. White's density matrix perturbation
(\ref{eq:DMP}) costs $O(2m^{3}wd^{3})$, but it can be reduced to that
of DMRG3S if the equivalence of the approaches is taken into
account. 

The Jacobi-Davidson correction (\ref{eq:Proj_JD}) is more expensive
than the Lanczos one because the GMRES solver required $K$ iterations,
scaling as $O(2m^{3}wK)$, similar to the diagonalization step (ii).
However, the missing factor $d$ can be used to compensate the greater
cost of the single-site diagonalization. The advantage is the splitting
into smaller problems, which typically decreases $K$.

Step (iii) can be replaced by  $\tilde{\psi}=\hat{H}\psi$, which mathematically brings us to the same new state Eq. (\ref{eq:enrich}). We find some cases where, starting from the compression of the exact $\tilde{\psi}=(H-E)\psi$, the sweeping $\langle\tilde{\psi}|\hat{H}|\psi\rangle$ setting $\tilde{\psi}=H\psi$ improves the energy of the compression at step (iv).

\section{Results}\label{sec:results}

In this section we present results for our algorithm for the Heisenberg spin $S$ chain and for free fermions. The first case is a standard benchmark in the literature~\cite{white2005density,hubig2015strictly,dolgov2015corrected}, and it will allow us to analyze the effects of the physical dimension $d$ by increasing the size of the spin representation, $d=2S+1$. The case of free fermions gives rise to a gapless system in the thermodynamic limit, and exhibits a situation where the entanglement is too large to be captured with a bounded MPS. We will also compare our results with those obtained using single-site DMRG (DMRG3S), finding similar or improved convergence.

\begin{figure}
\begin{centering}
\includegraphics[width=11.5cm,angle=270]{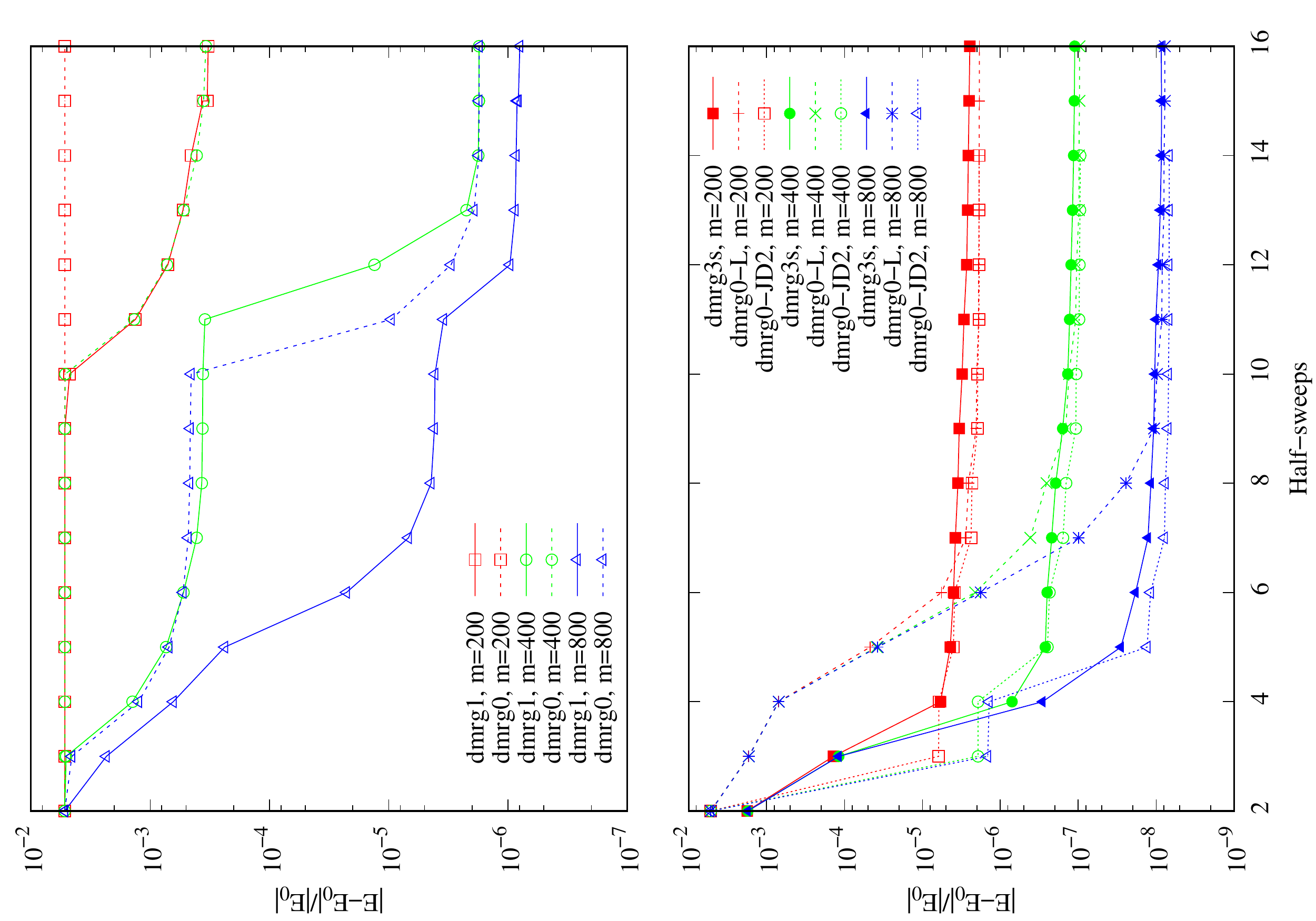}
\par\end{centering}
\caption{Normalized error vs sweeps of the ground state energy for the periodic
$S=1$ ($d=3$) Heisenberg spin chain, $L=100$. Top panel: DMRG0
compared to the single-site DMRG, both without enrichment. Bottom
panel: DMRG3S vs DMRG0 using one Lanczos or two Jacobi-Davidson corrections.
We take $E_{0}=-140.148404$ as the reference value~\cite{white2005density}.}\label{fig:s1}
\end{figure}

The Heisenberg Hamiltonian is
\begin{equation}
H=\sum_{i=1}^{L}\hat{S}_{i}\cdot\hat{S}_{i+1}\,,\label{eq:Hs1}
\end{equation}
and the site dimension corresponds to $d=2S+1$ . Fig.~\ref{fig:s1} shows the
convergence of the energy $E$ using DMRG0 and DMRG3S for $S=1$ and $L=100$. The top panel
presents the methods without enrichment. As expected, DMRG0 gets stuck
in a considerably greater energy than DMRG3S does for the same bond
dimension $m$. This is because DMRG0 updates only $m^{2}$ parameters
for each position $i$, compared to DMRG3S which updates $m^{2}d$.
In fact, the updates of DMRG0 do not cover the number of parameters
per site $m^{2}d$ of the MPS ansatz. 

On the other hand, the observed convergence ratios at the bottom panel of Fig.~\ref{fig:s1}
for DMRG0-L and DMRG-JD2 are surprising. Particularly, DMRG0-L makes
only $m^{2}$ updates per site at step (ii) enriched by a cheap direct
residual calculation in step (iii). DMRG-JD2 would cost in principle
like DMRG3S, although we remark that the former deals with $d$ problems
of $1/d$ smaller size.

\begin{figure}
\begin{centering}
\includegraphics[angle=270,width=8.5cm]{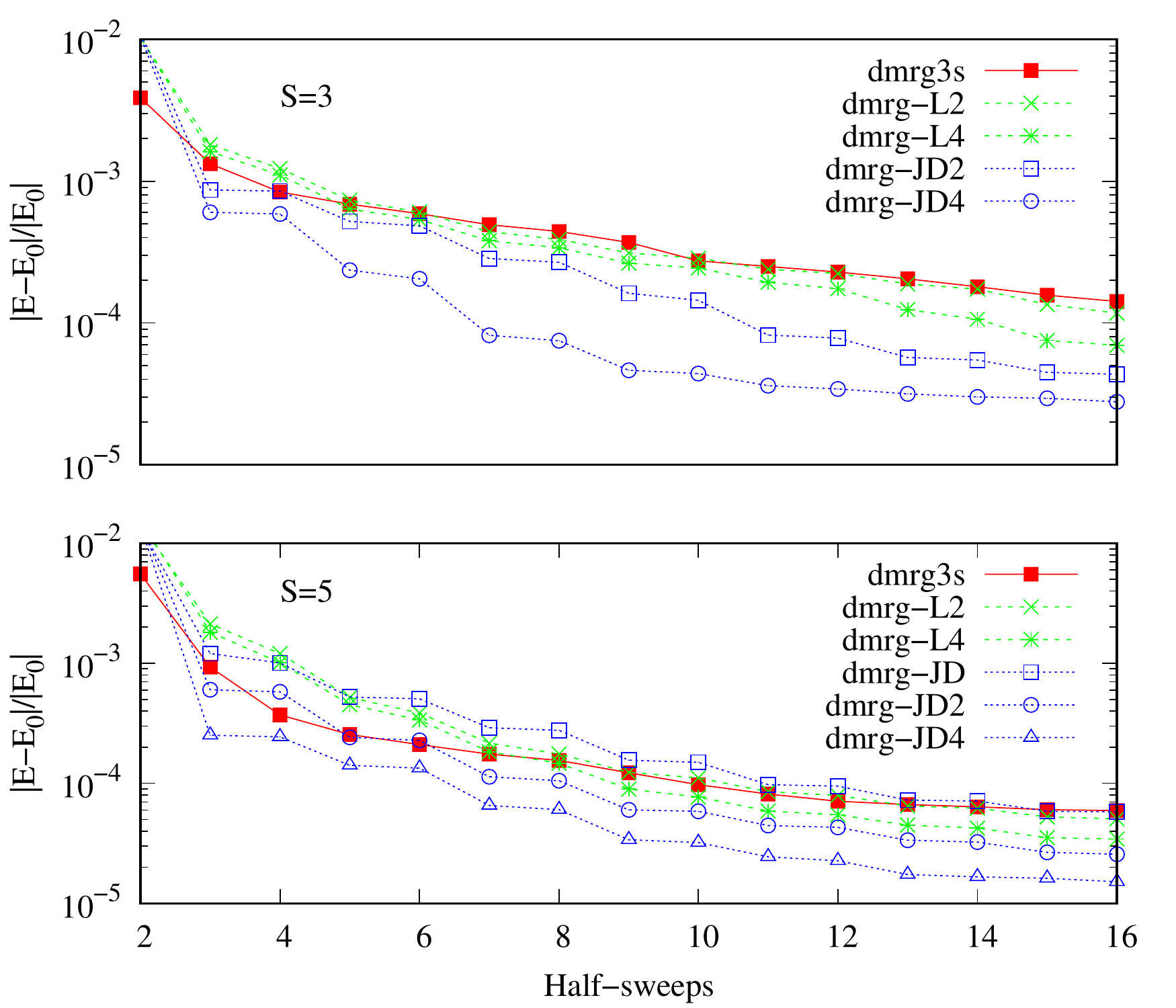}
\par\end{centering}
\caption{Same as Fig.~\ref{fig:s1} bottom panel with $S=3$ ($d=7$) and $S=5$ ($d=11$), $L=20$, $m=200$. We take
$E_{0}=-202.710579$ and $E_{0}=-537.28569$, respectively, obtained as the lowest eigenvalue of (\ref{eq:h_2x2mat}) with DMRG-JD4. Note that DMRG-JD4 uses a Hilbert space 5 times larger, and corresponds to $m \approx \sqrt{5} \times 200$.}\label{fig:s5}
\end{figure}

One of the important features of our method is that the optimization step (ii) does
not depend on $d$, which opens up the possibility of analyzing systems with large $d$.
Although we postpone a more detailed study of this aspect to a future work, let us briefly present results for the Heisenberg model
(\ref{eq:Hs1}) with $S=3$, namely
$d=2S+1=7$, and  $S=5$, i.e. $d=11$. The results
are shown in Fig.~\ref{fig:s5}. We find that both the Lanczos and the
Jacobi-Davidson corrections in DMRG0 outperform the convergence speed of DMRG3S.\footnote{We have checked that DMRG3S continues to decrease the energy if the sweeping continues, i.e. does not get trapped in a false minimum.} This is due to the fact that our global enrichment method is more powerful than the one used in~\cite{hubig2015strictly}. The approach in~\cite{hubig2015strictly} uses a homogeneous weight $\beta$ (\ref{eq:sub_exp}) for enrichment, while our approach is motivated by finding the optimal correction.

\begin{figure}
\begin{centering}
\includegraphics[angle=270,width=8.5cm]{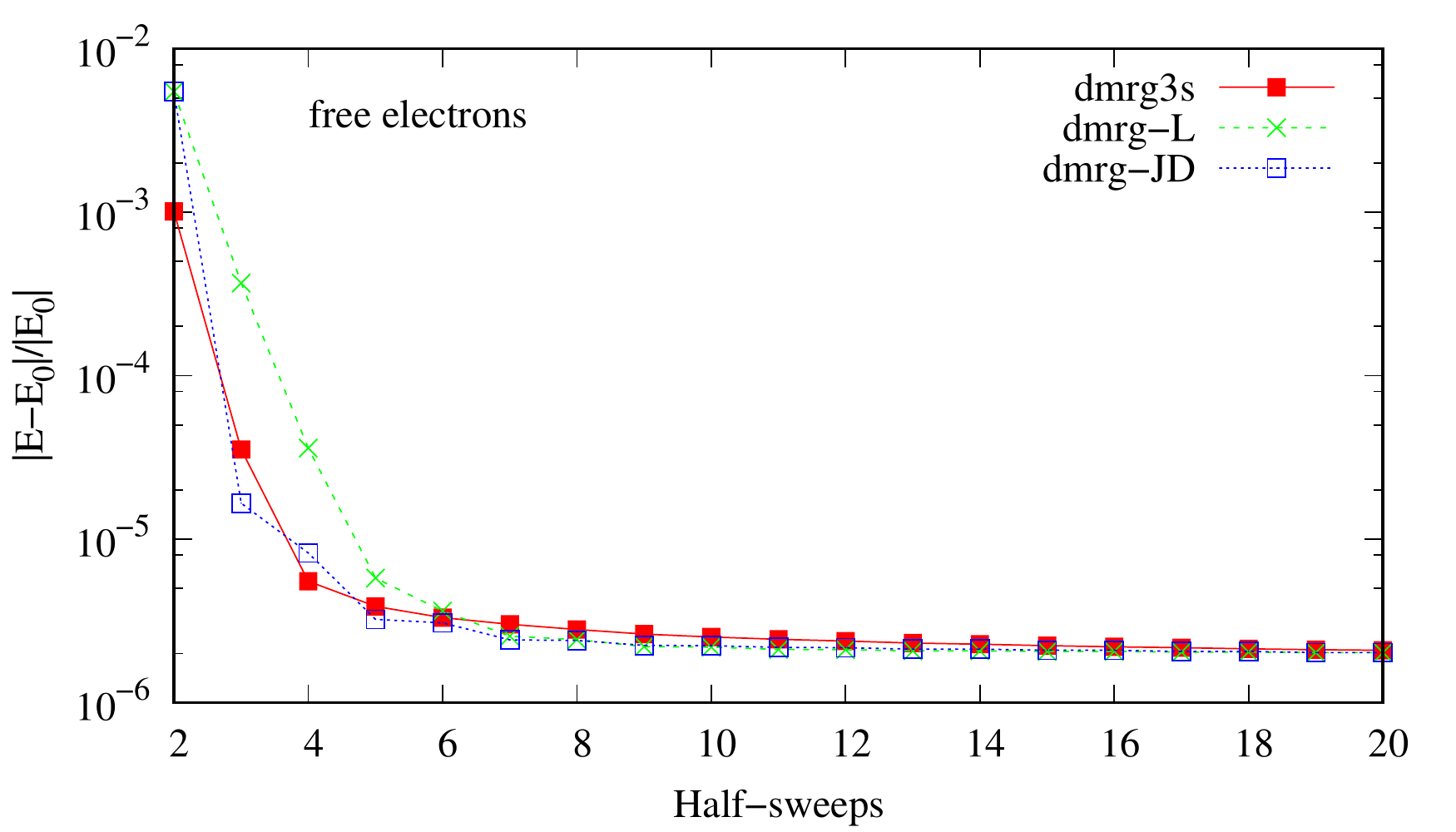}
\par\end{centering}
\caption{Comparison of DMRG3S and DMRG0 (Lanczos and Jacobi-Davidson) for free electrons in a one-dimensional periodic chain. The exact result for the ground state energy is $E_0=-63.6410319075$.}\label{fig:fermions}
\end{figure}

Finally, we study free fermions with periodic tight-binding Hamiltonian
\be
H =  \sum_{i} (c_{i+1}^\dag c_i+ c_i^\dag c_{i+1})\,,
\ee
where $c_i^\dag$ creates a fermion at site $i$, and $i = 1,\ldots, L$.
The system is gapless in the infinite size limit, with an entropy that grows like $S(R) \approx \frac{1}{3} \log R$, for a region with $R$ sites. For this reason, the system cannot be simulated with an MPS of fixed bond dimension, and the problem is quite challenging for DMRG. Fig.~\ref{fig:fermions} shows the ground state energy error for $L=100$. The problem of the growth in MPS dimension is not ameliorated by any optimization method. Capturing the logarithmic growth in the entropy requires changing the groundstate ansatz, for instance using MERA~\cite{PhysRevLett.101.110501}.

\section{Conclusions and perspectives}\label{sec:concl}

In this work, we have presented the zero-site DMRG, a new algorithm to find MPS ground
states, with the feature that the local optimization does not
depend on the site dimension. We have also proposed a new space enrichment method that avoids
local minima and speeds up the convergence ratios to the level of state-of-the-art
single-site algorithms. Conceptually, the local optimization of the wavefunction
and the renormalization and enrichment become separate steps.

Both the DMRG0 and the enrichment methods (Lanczos and Jacobi-Davidson)
open up the possibility of several developments and extensions. 
Since the optimization approach is independent of $d$ (the site physical dimension), DMRG0 could be well-suited to analyze systems with large $d$. This limit is interesting both theoretically as well as for its applications, such as in the Kondo lattice, dimensional reductions on cylinders, SYK-like non-Fermi liquids, holographic models, etc.

It would also be interesting to investigate in more detail the Lanczos and Jacobi-Davidson methods that we introduced. More nontrivial combinations of these approaches are possible. The Jacobi-Davidson corrections can also be applied to obtain excited states.
The global enrichment could be replaced
by a sequential local enrichment similar to~\cite{dolgov2015corrected,
dolgov2014alternating}. It would also be important to apply this to the single-site scheme, where we expect improvements over the local homogeneous enrichment methods used so far.

\acknowledgments

We would like to thank K. Hallberg for insights and constant encouragement, and K. Hallberg and D. Savostyanov for detailed feedback on the manuscript.
YNF and GT are supported by CONICET, UNCuyo, and CNEA. GT would like to acknowledge hospitality and support from the Aspen Center for Physics (NSF grant PHY-1607611, and Simons Foundation grant), and Stanford University, where part of this work was performed.

\bibliographystyle{apsrev4-1}
\bibliography{refs}

\end{document}